# Determining diffusivity and dissipation of diffuse body waves in concrete with multiple planar boundaries

Hao Cheng[a,*], Katrin Löer[b], Max A.N. Hendriks[a,c], Yuguang Yang[a]

[a] Department of Engineering Structures, Faculty of Civil Engineering and Geosciences, Delft University of Technology, 2628 CN Delft, the Netherlands
[b] Department of Geoscience and Engineering, Faculty of Civil Engineering and Geosciences, Delft University of Technology, 2628 CN Delft, the Netherlands
[c] Department of Structural Engineering, Faculty of Engineering, Norwegian University of Science and Technology, 7491 Trondheim, Norway
[*] Corresponding author: h.cheng-2@tudelft.nl

**Abstract**
This paper presents a new method for determining the diffusive properties of diffuse body waves, specifically diffusivity and dissipation, in concrete with multiple planar boundaries. Instead of relying on the analytical solution to the diffusion equation, which accounts for all possible reflections and has a significantly complex form, we propose a simpler method that considers the reflected energy originating from a limited number of image sources, making it easier to perform curve fitting and obtain diffusive properties. Our experimental findings highlight that distant image sources have a negligible effect on the acquired diffusive properties, suggesting that the number of image sources considered should be determined based on their maximum contribution to the main energy. Additionally, failing to account for the reflected energy during the fitting process results in an underestimation of both the diffusivity and dissipation. The obtained diffusive properties can be used to evaluate damage in concrete.

## 1. Introduction

Concrete is widely recognized as the predominant construction and building material worldwide. There has been a growing interest in non-destructive, real-time monitoring of the structural integrity of concrete structures, especially concrete infrastructures, over the past two decades [1-3]. Among the various techniques available, the ultrasonic approach has emerged as one of the most promising methods due to its ability to detect minor structural defects [4].

The ultrasonic technique can be classified into two categories: those relying on ballistic waves [5] and those relying on diffuse waves [6]. The ballistic wave-based method necessitates the use of a wavelength comparable to the length of the damage being detected in order to achieve high sensitivity [5], typically using frequencies in the range of several hundred kilohertz for early crack detection. However, high-frequency body waves experience significant attenuation in concrete, leading to a growing preference on the utilization of diffuse waves in concrete monitoring [7-9].

Diffuse waves, characterized by undergoing multiple scattering events, dominate in the coda segment of a wave signal [10]. Since diffuse waves in the coda have undergone significant scattering by heterogeneities in concrete, they travel a much longer path than ballistic waves and become highly sensitive to weak perturbations in the medium, such as stress changes [11, 12] and temperature changes [13, 14], due to the cumulative effect of these perturbations on travel time changes along the wave path.

Research on the application of diffuse waves in concrete structure monitoring can be broadly categorized into two groups: monitoring of a given region by exploiting diffusive properties of the whole region [15, 16], and localizing the disturbance using coda wave interferometry (CWI) in conjunction with the sensitivity kernel [17, 18]. Here, diffusive properties refer to diffusivity and dissipation, which describe the speed of energy transport and the amount of energy lost during that transport, respectively. The former research topic focuses on leveraging changes in diffusive properties to characterize variations in the micro- or meso-structure of the material. The latter research topic involves utilizing elastic waves within a specific time window and localizing disturbances in the sensor grid. The sensitivity kernels used in these research are generally constructed based on diffusive properties [19]. Therefore, both approaches rely on the accurate estimation of diffusive properties.

Diffusive properties of a medium can be extracted by fitting the wave energy profile with diffusion equations in one-dimensional [16, 20], two-dimensional [21, 22] or three-dimensional [15, 23-25] forms. However, it is important to note that the analytical solution of the diffusion equation is derived for an infinite medium. While concrete can be considered as an infinite medium in certain cases where the boundary is sufficiently far away, as observed in the literature [26, 27], the direct application of the diffusion equation is limited in most scenarios. To overcome this limitation, Ramamoorthy et al. [28] proposed an analytical solution of diffusion equation that incorporates the Neumann boundary condition. This modification, which considers all possible reflected energy, enables a theoretically more accurate estimation of diffusive properties within a two-dimensional rectangular domain [29-31] and three-dimensional cuboid domain [32, 33]. Nevertheless, the increased complexity in the expression of the diffusion equation makes curve fitting more intricate for obtaining diffusivity, rendering it less robust compared to the original diffusion equation in infinite media [32].

In this paper, we present a method to determine the diffusive properties of body waves in concrete members with planar boundaries. We treat the reflected energy as being generated by image sources, which is known as the image source method [19, 34-36]. However, applying the image source method to acquire diffusivity and dissipation of elastic waves in concrete with complex planar boundary conditions remains a challenge. Through the experimental observation in this paper, we find that it is not necessary to consider all possible reflections in the diffusion equation. Therefore, we adopt an improved diffusion equation by considering reflections originating from limited image sources, while maintaining the simplicity of the expression. This proposed method allows for accurate and efficient estimation of diffusion and dissipation while ensuring the robustness of the fitting process.

## 2. Theoretical background of diffusion equation

In a heterogeneous material like concrete, the wave transport regime is governed by the wavelength $\lambda$ and the average diameter of heterogeneities $\bar{d}$. The average diameter of heterogeneities is typically associated with the size of aggregates and air bubbles in concrete. Based on the different $\lambda$-$\bar{d}$ relation, the scattered elastic waves in concrete are generally categorized into three regimes [37, 38]:

- $\lambda \gg \bar{d}$ : when the wavelength is much greater than the average diameter of heterogeneities, the propagating wave encounters limited interactions with the heterogeneities. This regime is known as the Rayleigh regime or low-frequency limit. Born approximation [38], or its analogue, Rayleigh-Gans approximation [39, 40], is applicable in this regime.
- $\lambda \cong \bar{d}$ : when the wavelength is of a size that is comparable to the average diameter of heterogeneities, the wave has sufficient interaction with the heterogeneities. This regime is known as the stochastic regime. The Born approximation is approximately correct when the wavelength is comparable to the mean size of heterogeneities [38].
- $\lambda \ll \bar{d}$ : when the wavelength is much smaller than the average diameter of heterogeneities, the wave strongly interacts with the heterogeneities. This regime is known as the geometrical regime or high-frequency limit. In this regime, the Born approximation fails [40, 41], and Anderson localization [42, 43] might be observed.

The energy transport in the Rayleigh and stochastic regimes can be described using the radiative transfer equation (RTE) [44]. However, the complex nature of RTE presents challenges for its practical application in engineering. As an alternative, the diffusion equation [45] is commonly used when the energy transport occurs in the Rayleigh and stochastic regimes [46-48]. Such a simple scalar equation has been successfully and extensively utilized in describing the energy transport of various phenomena, including electromagnetic waves [49], acoustic waves [50] and elastic waves [51]. In this section, we will begin by presenting the expression of the diffusion equation for an infinite medium. Subsequently, we will extend the diffusion equation to incorporate reflections from planar boundaries.

### 2.1 Diffusion equation in an infinite medium

The governing equation for diffusion can be written in the following form [34, 52]:

$$\frac{\partial E(x_i,t)}{\partial t} = D\nabla^2 E(x_i,t) - \sigma E(x_i,t) + E_0 \quad , \tag{1}$$

where $t$ represents the time and $x_i$ are the spatial coordinates. In a three-dimensional Cartesian coordinate system, $i$ can be 1, 2 and 3. $E(x_i,t)$ is the transport energy at location $x_i$ and time $t$, and $E_0$ represents the deposited impulse energy at initial location at time $t = 0$ [34]. The parameters $D$ and $\sigma$ represent the diffusivity and dissipation, respectively, that describe the characteristics of energy transport.

The diffusivity quantifies how quickly the energy spreads within the medium [6]. In this context, the diffusivity can be viewed as a measure similar to the 'spreading velocity' of the diffusion halo [16]. It is noteworthy that the pure diffusion process, in the absence of dissipation, generally follows energy conservation [26]. The mechanism of dissipation in concrete is likely a combination of viscous dissipation and internal friction [53]. Unlike the energy-conserving nature of the pure diffusion process, the dissipation process does not follow the energy conservation.

In an infinite $n$-dimensional medium, the solution of Equation (1) is given following[34]:

$$E(r,t) = \frac{E_0}{(4\pi Dt)^{\frac{n}{2}}} e^{\frac{-r^2}{4Dt} - \sigma t}, \tag{2}$$

where $r$ represents the distance between source and receiver. The logarithmic form of Equation (2) is commonly employed in fitting the diffusion equation:

$$\ln(E(r,t)) = \ln(E_0) - \frac{n}{2}\ln(4\pi Dt) - \frac{r^2}{4Dt} - \sigma t, \tag{3}$$

Taking the partial derivative of Equation (3) with respect to $t$ gives:

$$\frac{\partial \ln(E(r,t))}{\partial t} = -\frac{n}{2t} + \frac{r^2}{4Dt^2} - \sigma. \tag{4}$$

As indicted in Equation (4), when the magnitude of $t$ is very small, the slope of the logarithm of the energy against time is positive and dominated by the term $r^2/4Dt^2$. When the magnitude of $t$ is large, the slope of the logarithm of the energy against time is negative and dominated by $\sigma$. One can also estimate the arrival time of the maximum energy from Equation (4) by setting the equation equal to zero and considering the fact that this arrival time must be positive:

$$t_{\text{maximum energy}} = \frac{-\frac{n}{2} + \sqrt{\frac{n^2}{4} + \frac{\sigma r^2}{D}}}{2\sigma}. \tag{5}$$

## 2.2 Diffusion equation in a medium with multiple planar boundaries

In the presence of complicated boundary conditions, the diffusion equation needs to be solved for the geometry of system [54], and this task presents a problem within the field of applied mathematics [19]. In situations where the boundaries of a concrete infrastructure are predominantly planar, the solution to the diffusion equation can be approximated by superimposing the energy from the real source with the additional energy reflected from the planar boundary. The reflected energy can be considered as originating from the image source [19, 34, 35], as illustrated in Figure 1. It is essential to underscore that the image source method is constrained to geometries formed by planar boundaries [55]. Considering that the acoustic impedance of concrete is around 20000 times higher than that of air [15], the transmitted energy from concrete to the air is in the order of $10^{-5}$, which is significantly smaller than the reflected energy. Therefore, we approximate the concrete-air boundary as the fully reflecting boundary. In this scenario, the energy can be represented as the summation of energy from the source and that from the image source [19]:

$$E_{\text{total}}(r,r',t) = E(r,t) + E(r',t). \tag{6}$$

where $r'$ represents the distance between image source and receiver.

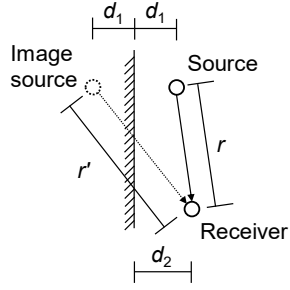

Figure 1. Schematic representation of energy superposition in the medium with a planar boundary.

After substituting Equation (2) into Equation (6), one can get the approximate expression of diffusion equation in a medium with one planar boundary:

$$E_{\text{total}}(r,r',t) = \frac{E_0}{(4\pi Dt)^{\frac{n}{2}}} e^{\frac{-r^2}{4Dt}-\sigma t}\left(1+e^{\frac{-\left[(r')^2-r^2\right]}{4Dt}}\right). \tag{7}$$

As shown in Equation (7), the contribution of an image source to the energy produced by the real source is calculated as $\exp\{-[(r')^2-r^2]/(4Dt)\}$ multiplied by the energy produced by the real source. In the case with multiple planar boundaries, the expression becomes:

$$E_{\text{total}}(r,r'_j,t) = \frac{E_0}{(4\pi Dt)^{\frac{n}{2}}} e^{\frac{-r^2}{4Dt}-\sigma t}\left(1+\sum_{j=1}^{N}e^{\frac{-\left[(r'_j)^2-r^2\right]}{4Dt}}\right), \tag{8}$$

where $N$ represents the number of image sources. The logarithmic form of Equation (8) is commonly employed in fitting the diffusion equation:

$$\ln\left(E_{\text{total}}(r,r'_j,t)\right) = \ln(E_0) - \frac{n}{2}\ln(4\pi Dt) - \frac{r^2}{4Dt} - \sigma t + \ln\left(1+\sum_{j=1}^{N}e^{\frac{-\left[(r'_j)^2-r^2\right]}{4Dt}}\right). \tag{9}$$

## 3. Concrete member, sensor layouts and measurements

In this paper, we use one concrete beam to demonstrate the method. This beam was chosen for the demonstration because the acquired diffusivity and dissipation values can be utilized to assess the damage condition of the beam during load testing. Additionally, these diffusive properties enable the localization of cracks during the initial stages of the load test using coda waves and the sensitivity kernel.

The prefabricated beam is constructed using an environmental-friendly cementitious material called alkali-activated (geopolymer) concrete [56]. The beam comprises two components: a prestressed beam with a height of 300 mm, along with a cast-in-situ layer atop the beam. The prestressed beam is manufactured in a prefabricated concrete plant. The cast-in-situ layer is applied 28 days after the beam casting. It is essential to note that the geopolymer concrete used in the topping layer is supplied by a commercial company and features a different composition compared to the mixture used for prestressed beams.

The piezoelectric sensors, namely smart aggregates (SAs), whose resonant frequency is around 80 kHz [57], are embedded inside this beam to serve as the actuators and receivers. These sensors are positioned at the mid-span in the beam. There are 12 SAs, designated from BB1 to BB12, as illustrated in Figure 2. The locations of each sensor in this beam can be found in Table 1. Measurements are taken between adjacent SAs within each row, where the SAs in the bottom row are not in direct communication with those in the top row. There are a total of ten SA pairs, and each SA pair involves two measurements, swapping the roles of transmitter and receiver between the first and second measurement. Hence, a total of 20 measurements are conducted. The SAs are situated with a minimum distance of 77 mm from the boundaries. Given that the frequency of interest is higher than 50 kHz and the Rayleigh wave velocity in concrete is approximately 2300 m/s [58], the maximum effective depth of penetration for Rayleigh waves is approximately 46 mm [59]. This depth is smaller than 77 mm, indicating that the contribution of Rayleigh wave-related energy transport in measurements can be neglected. During the data acquisition, each measurement involves gathering data by stacking five signals. The sampling rate for data acquisition is 3 MHz.

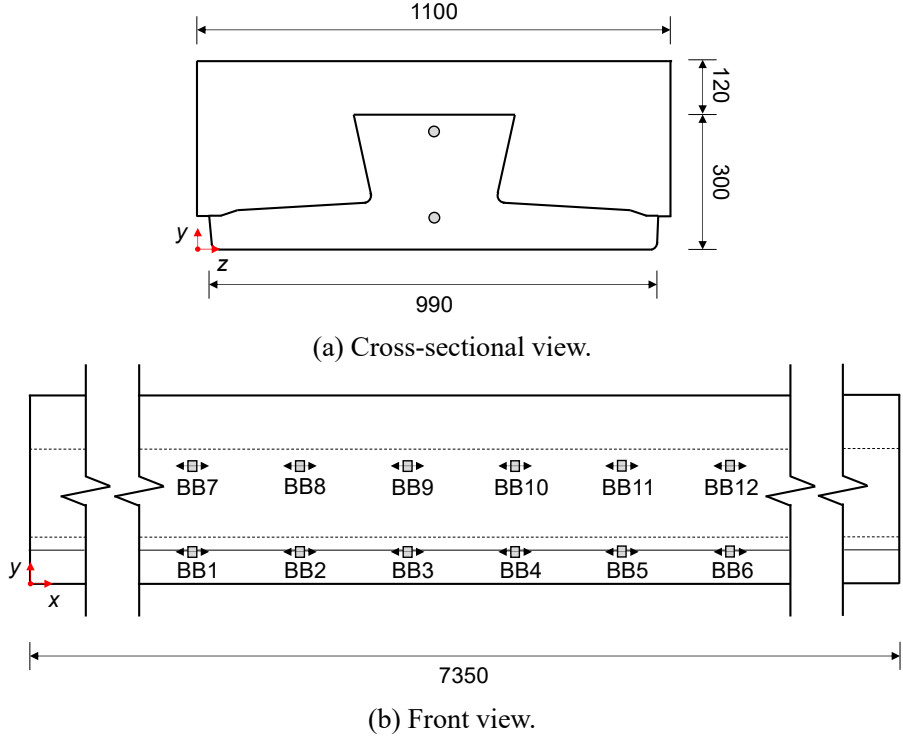

(a) Cross-sectional view.

(b) Front view.

Figure 2. Dimensions of the geopolymer concrete beam and the sensor layout (unit: mm; black arrow on the SA indicates the polarization direction of the sensor).

Table 1. Locations of SAs relative to the coordinates in Figure 2.

| Sensor | BB1 | BB2 | BB3 | BB4 | BB5 | BB6 | BB7 | BB8 | BB9 | BB10 | BB11 | BB12 |
|---|---|---|---|---|---|---|---|---|---|---|---|---|
| $x$ [mm] | 3070 | 3315 | 3560 | 3815 | 4070 | 4325 | 3070 | 3315 | 3560 | 3815 | 4070 | 4325 |
| $y$ [mm] | 77 | 77 | 77 | 77 | 77 | 77 | 277 | 277 | 277 | 277 | 277 | 277 |
| $z$ [mm] | 550 | 550 | 550 | 550 | 550 | 550 | 550 | 550 | 550 | 550 | 550 | 550 |

## 4 Diffusive properties of body waves in concrete beam
### 4.1 Methods for time-frequency analysis and curve fitting

Since diffusivity and dissipation are functions of frequency, these properties of body waves should be acquired for specific frequency bands. Here, the continuous wavelet transform (CWT) is used to transform the time-domain signal into a time-frequency spectrum. The analytical Morlet wavelet [60, 61] is selected as the mother wavelet. The filtering is performed in the frequency domain by multiplying the frequency spectrum of the received signal with a CWT filter bank $\Phi(\omega)$ consisting of $M$ Gaussian filters:

$$\Phi(f_i, \omega) = 2\mathrm{H}(\omega) e^{-\frac{[S(f_i)\omega - \omega_0]^2}{2}}, \quad (10a)$$

$$S(f_i) = \frac{\omega_0}{\pi} \frac{f_{\text{Nyquist}}}{f_i}, \quad i \in [1, M], \quad (10b)$$

where $\omega$, $f_{\text{Nyquist}}$ and $f_i$ represent the non-dimensional angular frequency, Nyquist frequency and centre frequency of the filter, respectively. H(·) and S($f_i$) are Heaviside step function and non-dimensional scale function. The parameter $\omega_0$ is a constant and is set to 6 to satisfy the admissibility condition, which requires the mother wavelet to be zero-mean [62]. As shown in Equation (10), the bandwidth of an individual filter is related to its central frequency. In both experiments, the range of central frequencies is selected from 20 kHz to 500 kHz with the frequency scale fineness of $2^{1/10}$. The products of CWT filter bank and received signals in the frequency-domain are then transformed back to the time-domain through the inverse fast Fourier transform (IFFT) to obtain the time-frequency spectrum.

Once the time-frequency spectrum is obtained, the wave energy is calculated by averaging the squared value of the amplitude within a specific time window with a selected length [34]. The length of the time window should be at least the reciprocal of the frequency of interest to minimize energy fluctuations caused by waveform oscillations. The centre time of the time window is taken as the elapsed time $t$ in Equation (9). When selecting the beginning and ending time windows for curve fitting, two factors need to be taken into consideration:
1. The beginning time window should encompass the first arrival of the wave to smoothen the wave energy, particularly to mitigate any sudden increases caused by the ballistic wave (i.e., Fig. 8 in Margerin et al. [39]).
2. The declining trend of the logarithmic energy should exhibit linearity or approximate linearity before the ending time window. This ensures the reliability of the fitted dissipation, as energy exhibits exponential decay when dissipation predominates.

### 4.2 Signal energy and parameters for fitting the diffusion equation
As mentioned in the previous section, the beginning and ending time windows for curve fitting need to be determined first for fitting the diffusion equation. To demonstrate how to determine these time windows, the logarithm of the ensemble-averaged energy at various frequency components, obtained from the signal received by BB2 with BB1 as the transducer, is plotted in Figure 3. Eight frequency components ranging from 50 kHz to 400 kHz, with an interval of 50 kHz, are selected. The length of time window is tentatively set as 40 μs (two times reciprocal of 50 kHz) for all frequency components, with an overlap of 20 μs between adjacent time windows. As shown in Figure 3, the ensemble-averaged energy of all frequency components initially experiences a rapid rise, followed by a subsequent decrease. The ascending part primarily corresponds to the spreading of energy in space, while the decreasing part is mainly attributed to dissipation. As the frequency of the wave increases, the energy decays at a faster rate. The energy of components with frequencies of 350 kHz and 400 kHz reaches the noise level at around 1000 μs. Similar decay trends can be observed in signals from other transducer-receiver pairs. Moreover, the energy in the high-frequency regime, particularly at 350 kHz and 400 kHz, exhibits drastic decay making the linear decrease associated with dissipation less pronounced. However, it can still be observed that the rate of energy reduction for 350 kHz and 400 kHz slows down after approximately 800 μs. Based on these observations, the centre time of the ending time window for curve fitting is set to 800 μs for all frequency components.

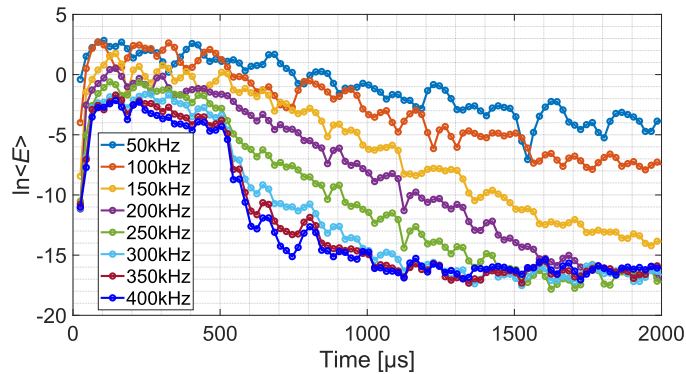

Figure 3. Logarithm of the ensemble-averaged energy (denoted as ln<$E$>) at various frequency components as a function of time received by BB2 while BB1 as the actuator.

As reported by Schubert and Koehler [27], the arrival time of the maximum energy density is dominated by diffusivity, while the subsequent decay is closely related to dissipation. However, as revealed in Figure 3, using time windows of equal length to capture the wave energy results in an imbalance of data points between the maximum energy arrival and energy decaying portions. For example, there will be more data points in the decaying portion (with a negative slope of the logarithm of the energy versus time) between approximately 200 μs to 800 μs, compared to the arising portion (with a positive slope). As indicted in Section 2.1, when the magnitude of $t$ is small, the slope of the logarithm of the energy against time is positive and dominated by the term $r^2/4Dt^2$, which is highly related to the diffusivity. When the magnitude of $t$ is large, the slope is negative and dominated by the dissipation $\sigma$. Therefore, the disparity in data sampling may cause the least-squares fitting to focus more on the decaying portion, where there are more data points, while neglecting the arising portion, which has fewer data points. This can lead to inaccurate estimation of the diffusion rate, which is more related to the arising portion.

To achieve this, the signal is divided into two parts: the initial part from 25 μs to 225 μs and the latter part from 225 μs to 725 μs. Starting from 25 μs can also help eliminate the influence of cross-talk on the obtained energy. The 1st time window, starting at 25 μs, has a length of 40 μs. The subsequent 10 time windows (2nd to 11th) also have a length of 40 μs, with each overlapping the previous window by 20 μs. The 12th time window, starting at 225 μs, has a length of 100 μs. The subsequent 10 time windows (13th to 22nd) have a length of 100 μs and overlap the previous window by 50 μs.

### 4.3 Influence of boundary reflections

Since the SAs are embedded at the midspan, far from the edges, there are four major types of reflections in the beam, illustrated using the SA pair BB1-BB2 in Figure 4. These reflections include: reflection from the bottom surface, reflection from the top surface, secondary reflections from bottom and top surfaces, and reflection from the front/back surface. Assuming that the diffusivity is 150 $m^2$/s and the maximum elapsed time is 800 μs, the maximum contributions of the reflected energy from these boundaries to the main energy, as shown in Figure 4, can be estimated using Equation (7): 95.2% from the bottom surface, 37.5% to the main energy from the top surface, 46.0% to the main energy from the secondary reflections, and 26.0% to the main energy from the front and back surfaces.

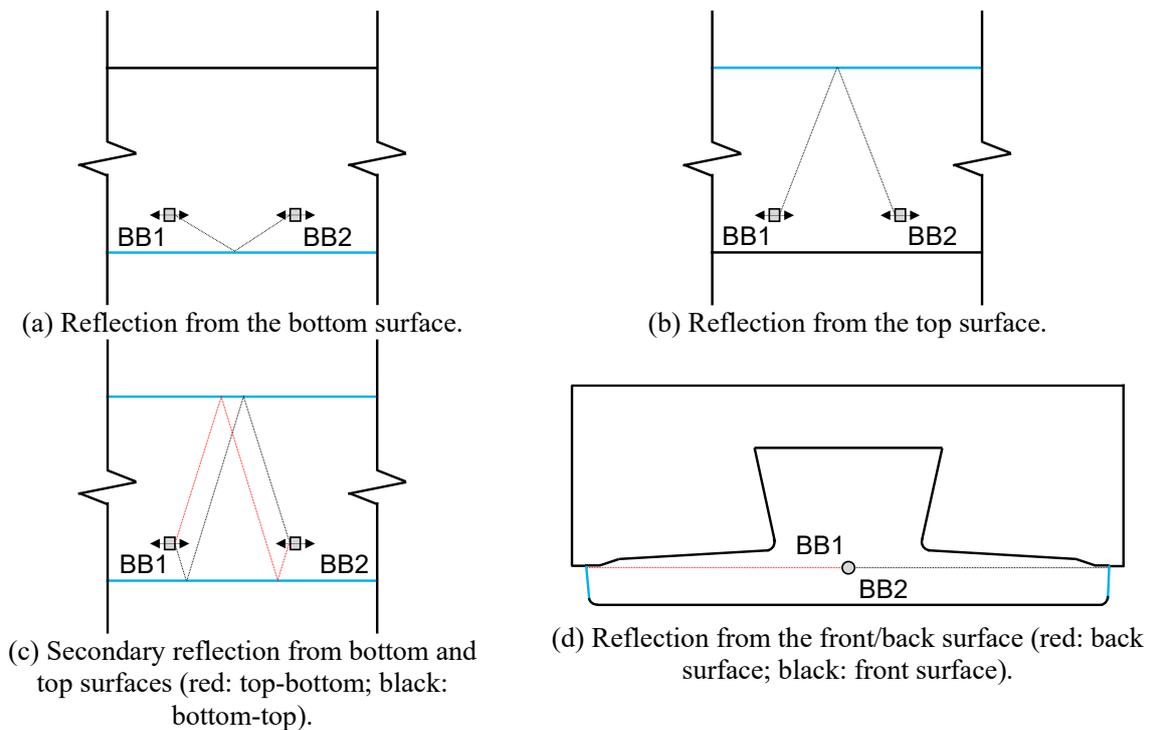

(a) Reflection from the bottom surface.

(b) Reflection from the top surface.

(c) Secondary reflection from bottom and top surfaces (red: top-bottom; black: bottom-top).

(d) Reflection from the front/back surface (red: back surface; black: front surface).

Figure 4. Illustration of four types of reflections from boundaries in the SA pair BB1-BB2. The reflecting boundaries are coloured in blue.

Figure 5 shows a typical experimental result of the signal energy of 200 kHz component from BB1-BB2 pair. The experimental result is fitted using both the diffusion equation in an infinite medium and the diffusion equation that accounts for reflected energy from top and bottom surfaces and secondary reflections. The curve fitting is evaluated using a least-square criterion to find the best-fit diffusivity and dissipation. Remarkably, although the forms of the diffusion equations used for fitting are different, the energy evolutions constructed using the fitted properties from diffusion equation in an infinite medium and the diffusion equation that accounts for reflected energy from top and bottom surfaces and secondary reflections are nearly identical. However, the acquired diffusivity and dissipation differ, as shown in Table 2.

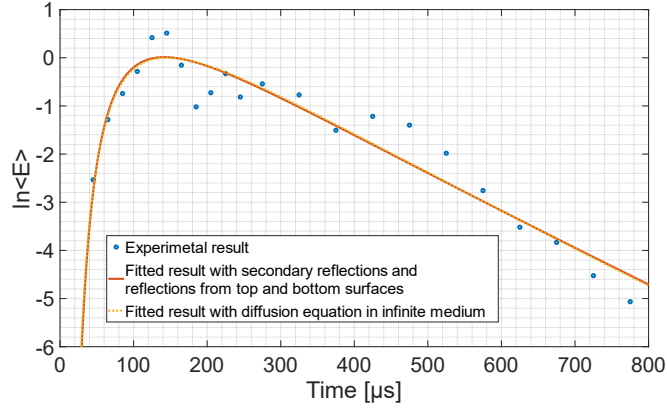

Figure 5. Comparison between fitted diffusion curves with and without considering boundary reflections (200 kHz component of the signal received by BB2 while BB1 as the actuator in the beam).

Table 2. Fitted results with different amount of image sources.

| Diffusion equation used for fitting | Fitted results | | |
|---|---|---|---|
| | $\ln E_0$ [-] | Diffusivity $D$ [m$^2$/s] | Dissipation $\sigma$ [s$^{-1}$] |
| Infinite medium (dotted line in Fig. 5) | -0.53 | 44.4 | 6205 |
| Considering reflections from bottom surface | -0.90 | 47.4 | 6465 |
| Considering reflections from bottom and top surfaces | -0.89 | 47.2 | 6505 |
| Considering reflections from bottom and top surfaces, and secondary reflections (solid line in Fig. 5) | -0.89 | 47.2 | 6523 |
| Considering reflections from bottom and top surfaces, secondary reflections, and reflections from front and back surfaces | -0.89 | 47.2 | 6525 |

Since a portion of the wave energy reflects from the boundary, the diffusion equation without considering boundaries interprets this reflected energy as part of the initial energy emitted by the source. Consequently, the calculated deposited energy ($\ln E_0$ in Table 2) is higher than when the reflected energy is taken into account. Additionally, if one fits the diffusion equation in such cases without considering the reflection, both diffusivity and dissipation will be underestimated, as indicated in Table 2.

Although the reflected energy has a non-negligible effect on the diffusivity and dissipation, it is unnecessary to account for all reflections. In scenarios presented in Table 2, considering reflections from the bottom and top surfaces, along with secondary reflections, already provides sufficiently accurate results. Consequently, only these image sources will be utilized in the fitting process for the bottom row of SAs in the beam.

The fitted diffusive properties as a function of frequencies, utilizing the mentioned reflections for the bottom row of SAs (BB1 to BB6), are depicted in Figure 6. In these figures, the box plot is employed to visually demonstrate the spread of diffusive properties. Each box represents the interquartile range (IQR) of the dataset, with the line inside indicating the median value. The top and bottom edges of the box represent the medians of the upper and lower halves of the dataset, respectively. Outliers, depicted as circles, are values that exceed 1.5 times the IQR above or below the top or bottom of the box. The lines extending from each box connect the median of the upper half to the maximum non-outlier data value and the median of the lower half to the minimum non-outlier data value. Similar to the observations in Table 2, neglecting reflections leads to consistently lower estimations of both diffusivity and dissipation. Moreover, the diffusivity of elastic waves in the low-frequency regime, characterized by faster energy spreading, is more affected by boundary reflections.

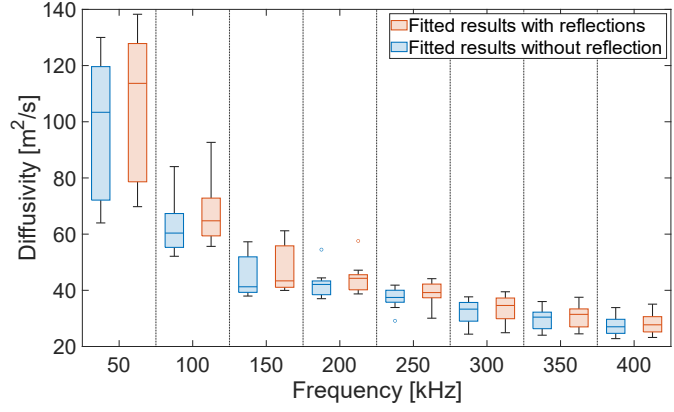

(a) Diffusivity.

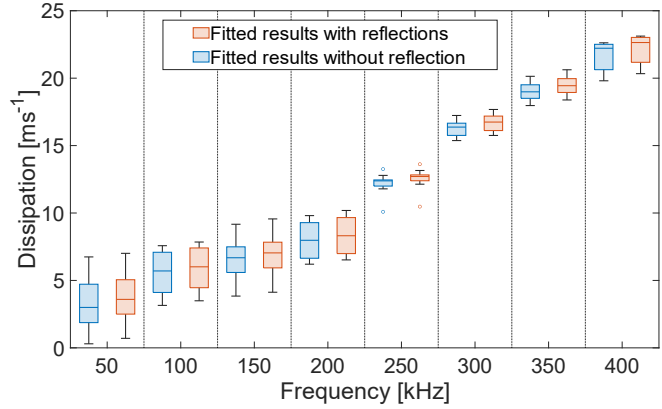

(b) Dissipation.

Figure 6. Comparison between the fitted diffusive properties with and without considering boundary reflections using signals from BB1 to BB6 in the beam.

### 4.4 Influence of concrete-concrete interface

As introduced in Section 3, the beam comprises two components: a prestressed beam and a cast-in-situ layer atop the beam. Since the cast-in-situ layer is applied 28 days after the beam casting, there might be a solid-solid interface between the prestressed beam and the top layer, which might affect diffusive properties of elastic waves in concrete. To investigate the influence of concrete-concrete interface on the diffusive properties, Figure 7 presents the diffusivity and dissipation obtained from both the top and bottom rows of SAs in the beam. It should be noted here that only reflections from solid-air boundaries are considered in the fitting process, assuming a perfect interface between the prestressed beam and the cast-in-situ layer. The fitted results reveal no apparent divergence in diffusivity between different SA rows, despite the small distance of 23 mm between the top row of SAs and the concrete-concrete interface. This suggests that the concrete-concrete interface has limited influence on diffusivity in this specimen. There are several outliers in the fitted results shown in Figure 7(a). Specifically, two data points from top row of SAs at 50 kHz show diffusivity values exceeding 200 $m^2$/s. These two outlier data points with extremely high diffusivity values will be excluded in the following analysis.

The wave energy shows a consistent faster decay in the region of the top row SAs, manifesting as a higher magnitude in dissipation. This observation can be attributed to three potential explanations. Firstly, the geopolymer concrete in the cast-in-situ topping layer may exhibit higher dissipation compared to the concrete in the prestressed beam, leading to a higher dissipation of elastic waves in this particular region. Secondly, it is possible that a portion of the wave energy leaks to the interface, which is also highlighted by Trégourès and van Tiggelen [63]. Thirdly, the region where the bottom SA row is embedded experiences higher compressive stress, which can lead to lower dissipation. This could happen when the internal friction dominates the dissipation process [53].

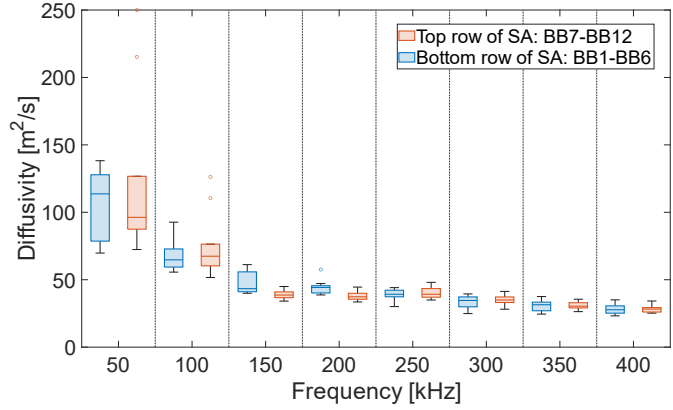

(a) Diffusivity.

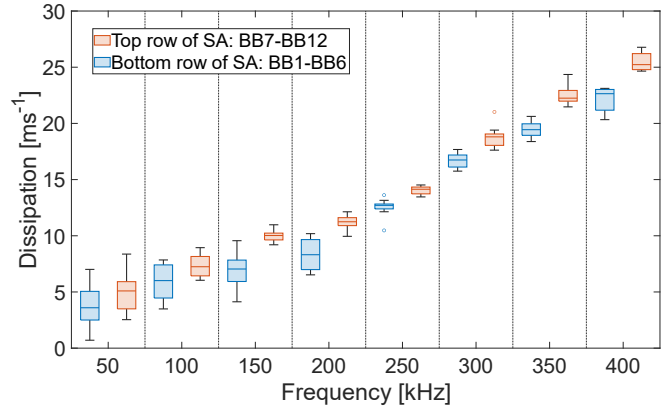

(b) Dissipation.

Figure 7. Comparison between the fitted diffusive properties from the top and bottom rows of SAs in the beam.

## 4.5 Summary statistics for diffusivity and dissipation in the beam

In Section 4.3 and 4.4, we discussed the influence of boundary reflections and the concrete-concrete interface on the diffusive properties in the beam, respectively. In this section, we present summary statistics for the diffusivity and dissipation of body waves in the beam, obtained using data from both the top and bottom rows of SAs, as shown in Figure 8. The diffusivity exhibits a decreasing trend with increasing wave frequency. The lower diffusivity observed at higher frequencies can be attributed to increased interactions between elastic waves and scatterers in concrete due to the shorter wavelength. The dissipation shows a linear increase with the wave frequency, which is in line with the result reported by Anugonda et al. [26]. In their work, the diffusion equation for an infinite medium is used to fit the diffusivity and dissipation, as the distant boundaries in their experiment have minimal impact on the diffusive properties.

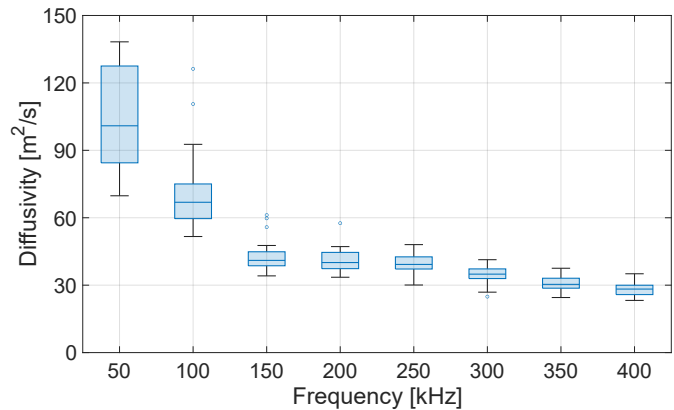

(a) Diffusivity.

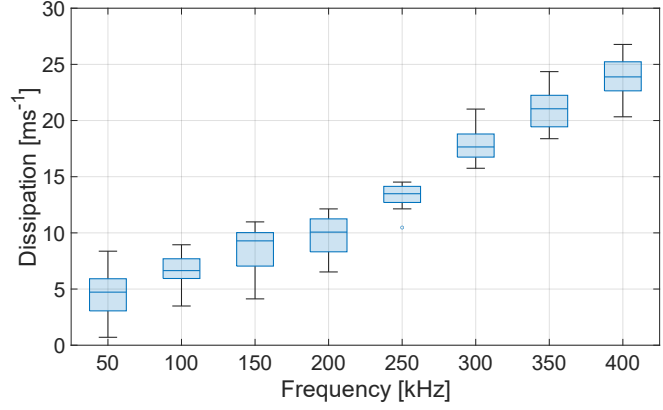

(b) Dissipation.
Figure 8. Diffusive properties of body waves in the beam acquired using all SAs taking into account reflections from the bottom and top surfaces, along with secondary reflections.

### 4.6 Arrival time of maximum energy

In this section, we will present the arrival time of maximum energy, which can be used to ensure that the fitted diffusive properties are reasonable. This time can be calculated using Equation (5) in conjunction with the diffusivity and dissipation given in Section 4.5. In practice, the arrival time of maximum energy cannot precede the first arrival of longitudinal waves. Therefore, any arrival time of maximum energy earlier than this threshold, approximately 55 μs as determined using the arrival time picker based on the Akaike Information Criterion [64], is deemed unreliable.

The arrival time of maximum energy as a function of frequencies is depicted in Figure 9, and the arrival time of longitudinal waves is highlighted as a red dotted line. By checking arrival times of maximum energy, the majority of diffusive properties obtained using the improved diffusion equation in geopolymer specimens is deemed reliable and trustworthy.

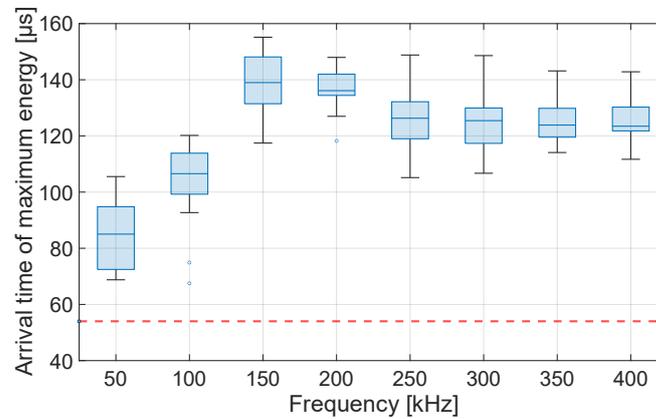

Figure 9. Arrival time of maximum energy in the beam (red dashed line indicates the mean arrival time of longitudinal waves).

### 5. Recommendations for applying the proposed method

Based on the results presented in this paper, we suggest the following steps for fitting the diffusivity and dissipation in concrete specimens with planar boundaries:
1. Filter the raw signals obtained from experiments to focus on the frequency of interest. In this paper, the CWT is utilized for this purpose, and the frequencies of interest are chosen from 50 kHz to 400 kHz with an 50 kHz interval.
2. Determine the ending time window based on the declining trend of the logarithmic energy, which should exhibit linearity or approximate linearity.
3. Determine the locations and length of time windows to ensure that data points in the energy arrival

and energy decaying portions are comparable. The length of time window should be at least the reciprocal of the frequency of interest, and certain overlapping is needed between two adjacent time windows to mitigate fluctuations between adjacent windows. In this paper, The 1st time window, starting at 25 μs, has a length of 40 μs. The subsequent 10 time windows (2nd to 11th) also have a length of 40 μs, with each overlapping the previous window by 20 μs. The 12th time window, starting at 225 μs, has a length of 100 μs. The subsequent 10 time windows (13th to 22nd) have a length of 100 μs and overlap the previous window by 50 μs.

4. Calculate the logarithm of ensemble-averaged energy in each time window.
5. Evaluate the maximum contribution of reflected energy from image sources by using $\exp\{-[(r')^2-r^2]/(4Dt)\}$ in Equation (7) with diffusivity value $D=150$ m$^2$/s. Please be aware that the diffusivity of ultrasonic waves in concrete is typically lower than 150 m$^2$/s. Therefore, this calculated contribution is generally overestimated than the actual contribution. The selection of time $t$ is discussed in Section 4.1, where the declining trend of the logarithmic energy should exhibit linearity or approximate linearity before time $t$. In this paper, $t$ is set to 800 μs for all frequency components under investigation. Image sources with low contribution can be neglected, and the threshold used in this paper is 0.30.
6. Construct the new diffusion equation using Equation (9).
7. Fit the constructed diffusion equation to the logarithm of ensemble-averaged energy to determine the diffusivity and dissipation. Given the significant fluctuations in diffusivity and dissipation, it is important to analyse their characteristics statistically, as individual observations may hold limited significance.
8. Verify the arrival time of maximum energy using Equation (5) to avoid anomalous results.

The potential applications of the acquired diffusive properties, according to previous research, include:
- Characterizing the properties of heterogeneity in concrete [21, 27, 65];
- Evaluating cracking damage in concrete [15, 22-25, 28, 30-32, 66, 67];
- Assessing the self-healing process of concrete [29, 33];
- Investigating water depercolation in cement paste [20];
- Evaluating thermal damage in concrete [16];
- Localizing damage in concrete structures [17, 68, 69].

## 6. Discussion

The proposed method in this paper is applicable to concrete members with different concrete mixtures. However, caution should be exercised to avoid using elastic waves with extremely high frequencies, where the wavelength is much shorter than the diameter of scatterers in concrete, which is equal to the diameter of coarse aggregates according to Ramaniraka et al. [16, 65]. Such conditions might lead to localization [70], where energy transport cannot be described using the diffusion equation [71]. Additionally, the proposed method is designed for body waves, which propagate three-dimensionally in the medium. Therefore, it is recommended to install the sensor inside the concrete structure, ensuring the distance from the boundary is at least greater than one wavelength of the Rayleigh wave [59] with the lowest frequency of interest. For surface-bonded sensors, the experimental setup must be carefully designed, as these sensors detect both surface and body waves. Decoupling the effects of these two types of waves during post-processing can be challenging. Furthermore, as discussed in Table 2, it is unnecessary to account for all reflections. Instead, only reflections with a sufficiently high contribution to the main energy at the maximum elapsed time should be considered. In this paper, the maximum elapsed time is 800 μs, and the contribution threshold is set at 0.3.

The quantification of the influence of solid-solid interfaces on diffusivity remains an open question. Previous studies [34, 72] have suggested that elastic diffusivity is related to the diffusivity of longitudinal and transverse waves, as well as the energy equilibration ratio. The diffusivity of different wave modes is further influenced by wave velocity and transport mean free path, which is inversely correlated with the total scattering cross-section [73]. When considering an energy-conserved system with constant wave velocities, the presence of solid-solid boundaries introduces additional scattering events, leading to a larger total scattering cross-section and a smaller transport mean free path. Consequently, both the diffusivities of longitudinal and transverse waves are expected to be smaller. Together with the fact that the energy equilibration ratio may also be affected by the solid-solid interface [63], we anticipate a decrease in elastic diffusivity when such interfaces are present. However, in our measurements, we did not observe a significant impact of solid-solid interfaces on elastic diffusivity. A possible explanation is that both mechanical properties and scattering properties of concrete in the top layer and prestressed beam are similar, which results in comparable diffusivities..

Another crucial aspect to consider in the application of the diffusion equation in concrete is its range of validity, which is also highlighted in the article by Seher et al. [30]. The underlying assumption when utilizing the diffusion equation is that the elastic waves operate within the diffusive regime. If the elastic wave does not fall within this regime, the energy transport should be described using the radiative transfer equation. In 1996, Ryzhik et al. [74] introduced an effective criterion for determining the range of validity of the diffusion approximation: the diffusion approximation is only valid if the energy of the longitudinal and transverse waves meet the equilibration. It is important to note that the equilibration referred here pertains to a global energy equilibration, the behaviour of the entire wave field rather than individual wave components. We do not address the validity of the diffusion approximation in this paper because investigating it requires examining the multiple scattering of body waves and the energy conversion during this process in concrete, which remains a fundamental question. This issue will be explored in future endeavours as part of ongoing research on this topic.

This paper proposes a method to estimate diffusivity and dissipation of elastic waves in concrete members with planar boundaries. As introduced in Section 1, there are two possible applications of estimated diffusivity and dissipation in concrete structure monitoring: monitoring of a given region by exploiting diffusive properties of the whole region [15, 16], and localizing the disturbance using CWI in conjunction with the sensitivity kernel [17, 18]. The former one focuses on leveraging changes in diffusivity and dissipation to characterize variations in the micro- or meso-structure of the material, while the latter one involves utilizing coda waves together with the sensitivity kernel constructed using diffusivity and dissipation to localize disturbances in the sensor grid. The further use of diffusivity and dissipation for damage detection and localization in the geopolymer beam during the load test will be presented in our future work.

## 7. Conclusion

This paper focuses on determining two diffusive properties of body waves, namely diffusivity and dissipation, in concrete structural members with multiple planar boundaries. In contrast to the analytical solution of the diffusion equation under Neumann boundary condition found in the literature, we employ a simple approach that considers the reflected energy as originating from the image source. The number of image sources to be considered is determined by estimating their maximum contribution to the main energy. The experimental results demonstrate that neglecting the reflected energy during the fitting process underestimates the diffusivity and dissipation. The proposed method can be applied to estimate the diffusivity and dissipation of body waves in concrete members with planar boundaries.